\documentclass{article}
\usepackage[letterpaper, margin=1in]{geometry}
\usepackage{xcolor}
\usepackage{authblk}
\usepackage[utf8]{inputenc}
\usepackage{graphicx}
\usepackage{setspace}
\newcommand\pubnumber{SLAC-PUB-17661}
\newcommand\pubdate{\today}
\newcommand\pubblock{\rightline{\begin{tabular}{l} \pubnumber\\
          \pubdate \end{tabular}}}

\def\Title#1{\begin{center} {\Large #1 } \end{center}}
\def\Author#1{\begin{center}{ \sc #1} \end{center}}

\def\CCC{C$^{3}$~}
\def\ee{e^+e^-}

\newcommand\snowmass{\begin{center}\rule[-0.2in]{\hsize}{0.01in}\\\rule{\hsize}{0.01in}\\
\vskip 0.1in Submitted to the  Proceedings of the US Community Study\\ 
on the Future of Particle Physics (Snowmass 2021)\\ 
\rule{\hsize}{0.01in}\\\rule[+0.2in]{\hsize}{0.01in} \end{center}}

\begin{document}
\begin{titlepage}

\snowmass
\pubblock
\Title{Strategy for Understanding the Higgs Physics: \\The Cool Copper Collider}

\Author{\textbf{Editors:}}

\Author{Sridhara Dasu$^{44}$, Emilio A. Nanni$^{36}$, Michael E. Peskin$^{36}$, Caterina Vernieri$^{36}$} 

\Author{\textbf{Contributors:}} 
\Author{Tim Barklow$^{36}$, Rainer Bartoldus$^{36}$, Pushpalatha C. Bhat$^{14}$, Kevin Black$^{44}$, Jim Brau$^{29}$, Martin Breidenbach$^{36}$
% ...
}

\noindent
% ...
$^{14}${Fermi National Accelerator Laboratory, $^{\dagger}$\& University of Chicago } \\
% ...
$^{29}${University of Oregon}\\
% ...
$^{36}${SLAC National Accelerator Laboratory $^{\dagger}$\& Stanford University}\\
% ...
$^{44}${University of Wisconsin, Madison}\\
% ...

\end{titlepage}

\begin{abstract}
A program to build a lepton-collider Higgs factory, to precisely measure the couplings of the Higgs boson to other particles, followed by a higher energy run to establish the Higgs self-coupling and expand the new physics reach, is widely recognized as a primary focus of modern particle physics. We propose a strategy that focuses on a new technology and preliminary estimates suggest that can lead to a compact,  affordable machine. New technology investigations will provide much needed enthusiasm for our field, resulting in trained workforce.  This cost-effective, compact design, with technologies useful for a broad range of other accelerator applications, could be realized as a project in the US. Its technology innovations, both in the accelerator and the detector, will offer unique and exciting opportunities to young scientists. Moreover, cost effective compact designs, broadly applicable to other fields of research, are more likely to obtain financial support from our funding agencies. 
\end{abstract}

\section{Introduction}

Future facilities to enable experimental enquiry of the five intertwined science drivers~\cite{P5}, identified by the particle physics community, have been studied extensively. Three of the five HEP science drivers identified by the community are: the usage of the Higgs boson as a new tool for discovery, the identification of the new physics of dark matter, and exploration of unknown new particles, interactions and physical principles. Progress in these areas will be enabled by new collider facilities, at high energies and high luminosities. 


A Higgs Factory would provide improved precision over the LHC, resolving Higgs properties 10 times better, and will enable a broad range of investigations across the fields of fundamental physics, including the mechanism of electroweak symmetry breaking, the origin of the masses and mixing of fundamental particles, the predominance of matter over antimatter, and the nature of dark matter. The cleaner $\ee$ environment aided by beam polarization available at linear colliders could become a sensitive probe to reveal more subtle phenomena.

The International Linear Collider~\cite{ILC} (ILC), based on superconducting RF technology, has the most advanced design. 
The science case for the electron-positron Higgs factory has been well developed by the ILC community. Further refinements of the physics case were also made by the Future Circular Collider community, using their plans for a circular electron-positron facility (FCC-ee)~\cite{FCC} and CEPC~\cite{CEPCStudyGroup:2018ghi}. As opposed to linear machines, circular colliders are strongly limited by synchrotron radiation above 350–400 GeV center of mass energy. Moreover a linear collider allows for trigger-less operation, which could also open up to new physics signatures that are going undetected at the LHC.

The international community, represented by the ICFA, has expressed interest in the ILC as a global project, and the ILC is now under consideration for construction in Japan. 
However, for a long time now, Japan has not initiated a process to host this collider. In view of this, it is appropriate to consider other options.

The FCC-ee machine is pipe-lined 7 years after the end of the HL-LHC program and expected to start in 2048. It will require a large upfront civil engineering cost to build a 100-km tunnel. The ultimate goal of a proton machine (FCC-hh), which starts after the FCC-ee, is also acknowledged as attractive. 
Yet again, it is prudent to investigate alternate plans based on technologies which could enable compact designs and possibly provide a roadmap to extend the energy reach at future colliders.

The goals for a worthwhile alternative plan are two-fold: a) a near-term, cost-effective Higgs factory that could fit within an existing laboratory site and b) a longer-term prospect for accessing higher energies in lepton collisions, again in a cost-effective fashion.  

The Cool Copper Collider (\CCC)~\cite{cccwhitepaper} is a relatively new proposal to build a Higgs Factory with a 250 GeV energy collision energy based on a technology that offers the option for an adiabatic upgrade to 550 GeV, and possible extension to the TeV-scale. Beam parameters for \CCC-250 and \CCC-550 are given in Table~\ref{tab:cccpar}. 

\begin{table}[h!]
\begin{center}
\begin{tabular}{c c  c } 
   \hline
   CM Energy [GeV] & 250 & 550 \\
     Luminosity [x10$^{34}$/cm$^2$s] & 1.3 & 2.4 \\
Gradient [MeV/m] & 70 & 120 \\
Effective Gradient [MeV/m] & 63 & 108 \\   %

Length [km] & 8 & 8 \\

  Num. Bunches per Train  & 133 & 75 \\
  Train Rep. Rate [Hz] & 120 & 120 \\
  Bunch Spacing [ns] & 5.26 &  3.5 \\
  Bunch Charge [nC]  & 1 & 1 \\
    Crossing Angle [rad] & 0.014 &  0.014\\

  Site Power [MW] & $\sim$150 & $\sim$175 \\ 
 
\hline
 \end{tabular}
 \caption{Beam parameters for \CCC. The final focus parameters are preliminary.\label{tab:cccpar}}
\end{center}
\end{table}

While several competing technologies for electron-positron colliders exist for a Higgs Factory, only high accelerating gradient linear machines are likely affordable to get to the TeV-scale. We will then probe the Higgs self-coupling and top-Yukawa coupling to a precision that sheds light on a broad class of puzzles around the Higgs boson. Such a TeV-scale upgrade also opens the gateway to access to TeV-scale new physics, especially in unexplored/underexplored weakly coupled states that may well escape detection at the HL-LHC. New physics of this kind are well-motivated by general theoretical considerations, such as supersymmetry, compositeness considerations, generic models with hidden dynamics, and dark matter.
\CCC can also probe a broad range of TeV-scale new physics, potentially explaining the g-2 and flavor physics anomalies. 

Broader impacts of \CCC technologies are also important to consider. Its high-gradient technology will enable compact electron and X-ray photon sources, which are in high demand for medical, materials science, and other applications. An invigorated community of particle and accelerator physicists, pushing the limits of technology for particle acceleration, detection, measurement and analysis, are likely to attract and train a strong workforce.

The \CCC concept is thus an attractive strategy to address our community's need for an $\ee$ Higgs factory.  The purpose of this document is to encourage our community to support R\&D and participate in defining the emerging \CCC option as part of the Snowmass community study process.

\section{The \CCC Concept}

\CCC is based on a fundamental study and optimization of electromagnetic cavities for high accelerating fields on axis (the gradient) and low breakdown rates. This optimization yields a design with an iris too small to propagate the fundamental cavity mode. This is solved by a second discovery of a distributed manifold supply to each cavity separately, with proper phase and proper fraction of the RF supply through the wave-guide coupling. The resulting structure, although far too complex for tradition assembly techniques, is straightforwardly built from about two meter long Cu slabs machined by numerically controlled milling machines. It is noteworthy that \CCC could not have been realized without modern supercomputers for the RF design and modern NC machining techniques.  

The linac is cooled to $\approx 80$~K by liquid nitrogen to reduce the RF power requirements, and increase the acceleration gradient, upwards of 150 MeV/m~\cite{nasr2021experimental,bane2018advanced}. Thus, the acceleration gradient of the \CCC linacs~\cite{cccwhitepaper} is an order of magnitude increase over the SLC, a factor 4 over that of the ILC\cite{ILC}, and a factor of two over the normal-conducting NLC design~\cite{NLC}. The \CCC plans to reuse the final focus design of the ILC, which is optimized for up to 1 TeV operation, recouping much of the progress made in its design.

The proposed \CCC has an 8 km footprint that can reach 250 GeV center of mass energy using innovative technologies, with the possibility of a relatively inexpensive upgrade to 550 GeV on the same footprint, adding only more RF sources.

The linac is constructed from 9~m cryomodules, each of which houses eight 1~m distributed coupling, accelerating structures supported on four 2~m rafts. Each raft supports two 1 m accelerator sections and a quadrupole with a mechanically integrated Beam Position Monitor, coarse and fine alignment movers, and position sensor fixtures. The cryomodule has four penetrations for RF power, each with two waveguides, with each waveguide powering one accelerating structure. The total cryogenic thermal load for the complex at 250~GeV is 9~MW. This thermal load is removed through nucleate boiling of liquid nitrogen.  Liquid nitrogen flows by gravity through the cryomodule to replace the nitrogen that has evaporated. Thus, the linac must be horizontal with straight segments of about 1~km. Nitrogen gas is removed from the linac at penetrations that are spaced at approximately 1~km and transported to a re-liquification plant before being reintroduced into the linac.

Great effort has been expended towards the design and optimization of the accelerator complex for earlier linear collider designs. This excellent work can be leveraged to understand the pre-conceptual layout for the full \CCC accelerator complex. In particular, the ILC designs for the electron and positron sources, the damping rings, and the beam delivery system can be taken over directly for \CCC, with small changes to accommodate the \CCC beam format. As the design matures, these systems will be revised and further optimized. 

The baseline electron (polarized) and positron (unpolarized) sources are conventional LC designs. For the electron source, this consists of a polarized DC gun, buncher, and accelerator. However, we are also exploring the possibility of a polarized RF gun. For the positron source, the closest design relevant to the \CCC bunch structure is the CLIC design~\cite{linssen2012physics} (see sec. 3.1.3.2). Positron polarization is a possible upgrade once the full RF system is installed. An additional electron bunch train would be extracted from the Main Linac at high energy (125 GeV) and sent to an undulator for polarized photon production and transport to a positron production target. The positron beam must be transported to a damping ring before being accelerated by the Main Linac.

For the damping rings, a conservative design has two damping rings, one for the electrons and one for the positrons. A pre-damping ring is also utilized for the positron beam. Four electron and positron bunch trains are stored in each of the damping rings. The main damping rings have a $\sim$900~m circumference. The electron damping ring might be eliminated with the advent of a polarized RF photo-injector. 

Scaling the beam delivery system (BDS) for a maximum single beam energy of 275~GeV from 500~GeV for the ILC TDR~\cite{Bambade:2019fyw} reduces the length of the BDS by 500~m. Furthermore, we also remove the upstream polarimeter in favor of the downstream polarimeter reducing the length by an additional 200~m.  The assumption of a single downstream polarimeter will be weighed against the benefits of including an upstream polarimeter during the CDR preparatory phase. For \CCC, the downstream polarimeter can measure the polarization and the depolarization from beam-beam interactions by comparing interacting and separated beams. The length for the BDS is approximately 1.5~km on each side. Preliminary simulations of for 250~GeV CM indicate that a 1.2~km BDS is feasible for that energy~\cite{cccwhitepaper}.

\section{The \CCC Higgs Factory at 250 GeV}

The planned High Luminosity era of the LHC (HL-LHC), starting in 2029\footnote{This refers to the updated schedule presented in January 2022~\cite{LHCschedule}. As the LHC schedule is evolving, the starting date of HL-LHC could change.}, will extend the LHC dataset by a factor 10 allowing an increase in the precision for most of the Higgs boson couplings measurements~\cite{cepeda2019higgs}. Studies based on the 3000 fb$^{-1}$ HL-LHC dataset estimate that we could achieve 2-4\% precision on the couplings to $W$, $Z$, and third generation fermions. But the couplings to first and second generation fermions will still largely not be accessible and the self-coupling will only be probed with O(50\%) precision. 

There are good reasons to measure the Higgs boson properties at higher precision than will be possible at the HL-LHC.   With the basic Higgs mechanism for mass generation now demonstrated, the next task for Higgs studies is to search for the influence of new interactions that can explain why the Higgs field has the properties required in the SM. If the new particles associated with these interactions are too heavy to be produced at the HL-LHC, they can still make an imprint on the pattern of Higgs boson couplings.  If we wish to prove the existence of these effects and to understand their pattern, we will need a very precise understanding of the Higgs boson properties, with measurements of 1\% or better.  Through global analyses, such high precision measurements can also improve the interpretation of LHC data and lead to a stronger comprehensive map of where new physics might lie. 

An $\ee$ Higgs factory has a distinctive duty to gain insight on the Higgs Yukawa couplings at the next level beyond the third generation fermions. In the SM, the Higgs Yukawa couplings are exactly proportional to mass.  This simplistic picture deserves close scrutiny. Tagging of charm and strange quarks, as previously demonstrated at SLC/LEP,  gives effective probes for advancing this program. There is a broader program to investigate the potential deeper connection of Higgs boson with flavor and CP violation. 

\CCC follows a program very similar to that proposed for the ILC~\cite{Bambade:2019fyw}.  We thus expect similar results, subject to considerations described below, for similar integrated luminosities and detector capabilities~\cite{cccwhitepaper}. For \CCC we assume a conventional positron source with zero polarization, as opposed to 30\% for ILC. There is almost no difference between the two cases in the expected precision in Higgs boson couplings for a given luminosity~\cite{Fujii:2018mli}. Positron polarization does allow the collection of additional datasets that may be useful in controlling systematic errors.  \CCC is compatible with the addition of a polarized positron source as an upgrade.  For energies well above 500 GeV, the enhanced cross section for $e^-_Le^+_R$ collisions makes this advantageous. 

\CCC is expected to be cost-effective for reaching the energy of 500-600~GeV.  But its key point is that it provides a more secure basis for extension to higher energies.

\section{Upgrade to 550 GeV}

Although most of the gain in precision in Higgs boson couplings will be realized already at the 250~GeV stage, there are important reasons to continue the $\ee$ program to an energy of 500—600~GeV. First, this energy is above the crossover point at which the $WW$ fusion reaction $\ee\to \nu\bar\nu h$ overtakes  the $\ee\to Zh$ reaction and becomes the dominant mode of Higgs boson production.   This means that, in  the 550~GeV data, the Higgs boson is mainly produced by a different mechanism with different sources of systematic errors.  The \CCC-550 then additional dataset will provide independent channels for the Higgs coupling measurements. The analysis of the measurements at two different energies will provide a more robust interpretation of the Higgs couplings by also breaking degeneracies. This would be key to interpret possible deviations from the SM predictions observed at 250 GeV that then can be cross-checked in this 550 GeV dataset. The  $\sigma\cdot BR$ for $WW$ fusion to $h$ with decay to $WW^*$ depends quartically on the Higgs-$W$ couplings and thus is a very powerful addition to the dataset. Second, this energy is needed to give a substantial cross section for the process $\ee\to Zhh$, which allows us to measure the Higgs self-coupling. The expected 20\% error on the Higgs self-coupling will allow us to exclude or demonstrate at 5~$\sigma$ the large enhancements to the self-coupling typically needed in models of electroweak baryogenesis~\cite{DiMicco:2019ngk}. Third, this energy choice is well above the threshold for $\ee\to t\bar t$, far enough that top quark pairs are produced with relativistic velocities. 
For the measurement of the top quark Yukawa coupling through $e^+e^-\to t \bar{t} h$, it is even important to go above 500~GeV.
The statistical error on this measurement decreases by a factor 2 when one chooses 550~GeV rather than 500~GeV as the CM energy~\cite{Fujii:2015jha}, as the $\ee\to t \bar t h$ cross section rapidly increases as a function of $\sqrt{s}$.  In principle, higher energy gives more of an advantage, but this must be balanced against increased cost and footprint.  For the purpose of this paper, we have set the design energy of \CCC at 550~GeV.
A crucial difference between models in which the Higgs boson is elementary and those in which the Higgs are composite is that, in the latter class, the top quark is also partially composite, with modified $W$ and $Z$ couplings. The 550~GeV measurements have great power to test for this property. Indeed, at 500-600 GeV, the top quarks become relativistic and it is then possible to separate the various independent form factors.   Beam polarization is also very important for these measurements~\cite{Schmidt:1994bq}.  Moreover, at 550 GeV, an $\ee$ collider is as powerful as HL-LHC in searches for $Z’$ bosons and light quark and lepton compositeness and provides a complementary set of tests~\cite{fujii2019tests}.

\section{Upgrade to TeV-scale} 

The \CCC linacs can potentially be lengthened to increase the center of mass energy to about 1 TeV. Prior planning of the site for length extension and possible relocation of upstream components is necessary. Assuming one can get to this TeV-scale, the potential for $\ee \to \nu \bar{\nu}$ HH measurement becomes possible with an integrated luminosity of a few ab$^{-1}$ ~\cite{Barklow:2017awn}. 
New and exciting physics opportunities across the board are enabled with a TeV-scale upgrade. The  Higgs precision enabled by the 550 GeV run will be further improved, particularly the Higgs trilinear and top-Higgs Yukawa couplings. 
These precision coupling improvements allow us to probe compositeness scale well beyond TeV, which is outside the realm of the LHC probes~\cite{deBlas:2018mhx}.
Weakly coupled new physics could be buried under the QCD background from a hadron collider environment. The TeV-scale \CCC will provide definite answers and insights about new physics within the kinematic threshold. For instance, weak scale lepton partners (such as sleptons) and electroweak states (such as electroweakinos) that are well motivated can be discovered at \CCC. In particular, various recent hints from experimental data, such as the muon g-2 anomaly~\cite{Athron:2021iuf} and lepton-flavor-universality-violations in $B$ meson systems~\cite{Buttazzo:2017ixm}, all have a large fraction, if not the entire, of motivated and allowed regions at the TeV-scale. The TeV upgrade of \CCC can pair-produce these states and detect them or access them through the precision measurement on their loop-induced effects. Dark matter candidates, or more general, missing energy, can be revealed through missing energy searches~\cite{Han:2020uak}. Other exotic signatures, such as long-lived particles or dark showers, can also be probed at the upgrade \CCC~\cite{deBlas:2018mhx}. TeV-scale lepton colliders share a lot of exciting physics opportunities and potential; we can see them in reviews, e.g.~\cite{EuropeanStrategyforParticlePhysicsPreparatoryGroup:2019qin,AlAli:2021let}. 

\section{Site Options}

A \CCC $\ee$ collider could in principle be sited anywhere in the world. Projects of its magnitude will be presented and reviewed in international fora, and a community decision will be made regarding the actual site selection. Nevertheless, we note that the \CCC offers a unique opportunity to realize an affordable energy frontier facility in the US in the near term. We note that the entire \CCC program could be sited within the existing US national laboratories. 

\begin{figure}[h!]
    \centering
    \includegraphics[width=0.6\textwidth]{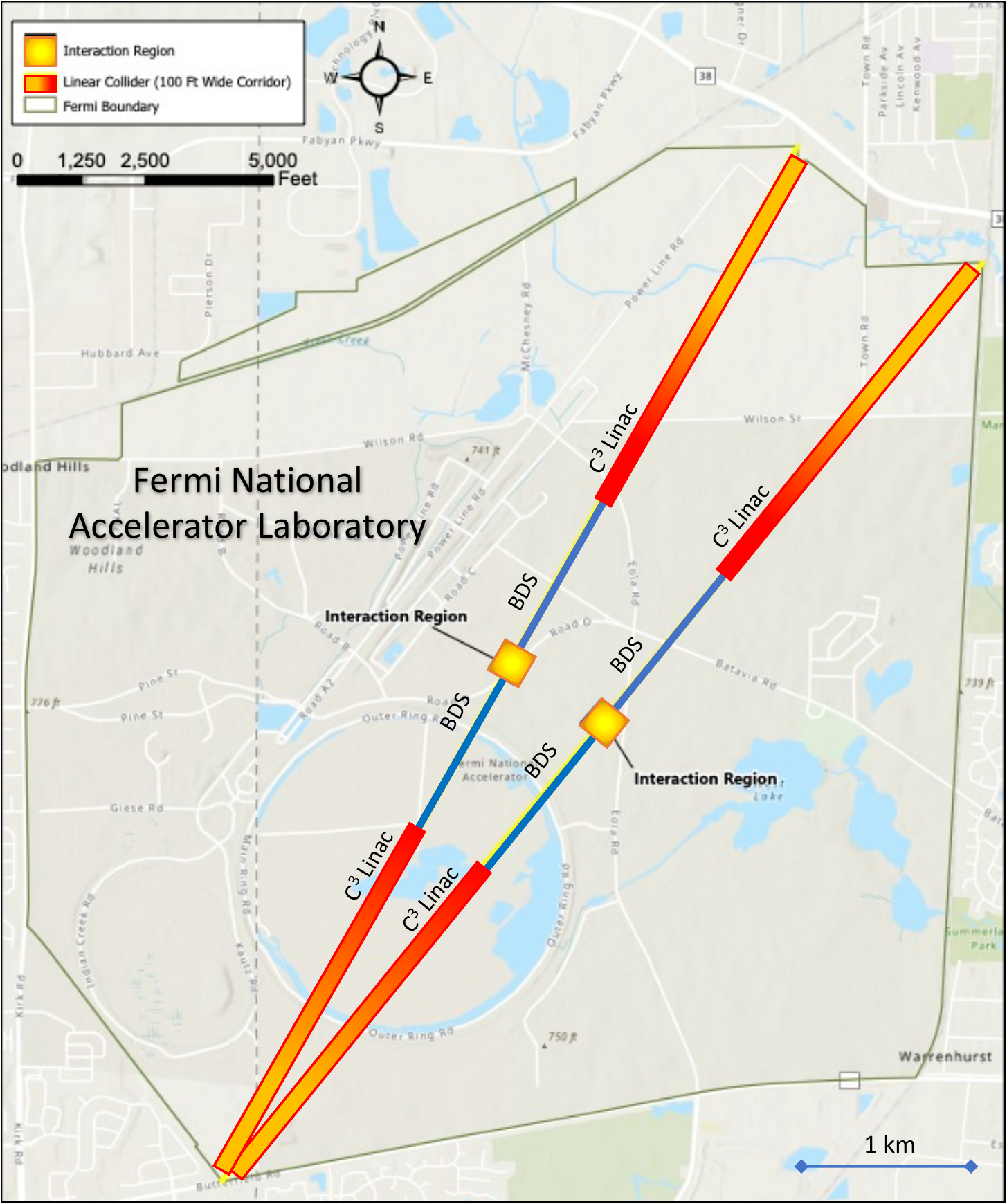}
    \caption{Possible locations for a 7-km footprint \CCC linear collider on Fermilab site.}
    \label{fig:LC_7kmLayout}
\end{figure}

\begin{figure}[h!]
    \centering
    \includegraphics[width=0.7\textwidth]{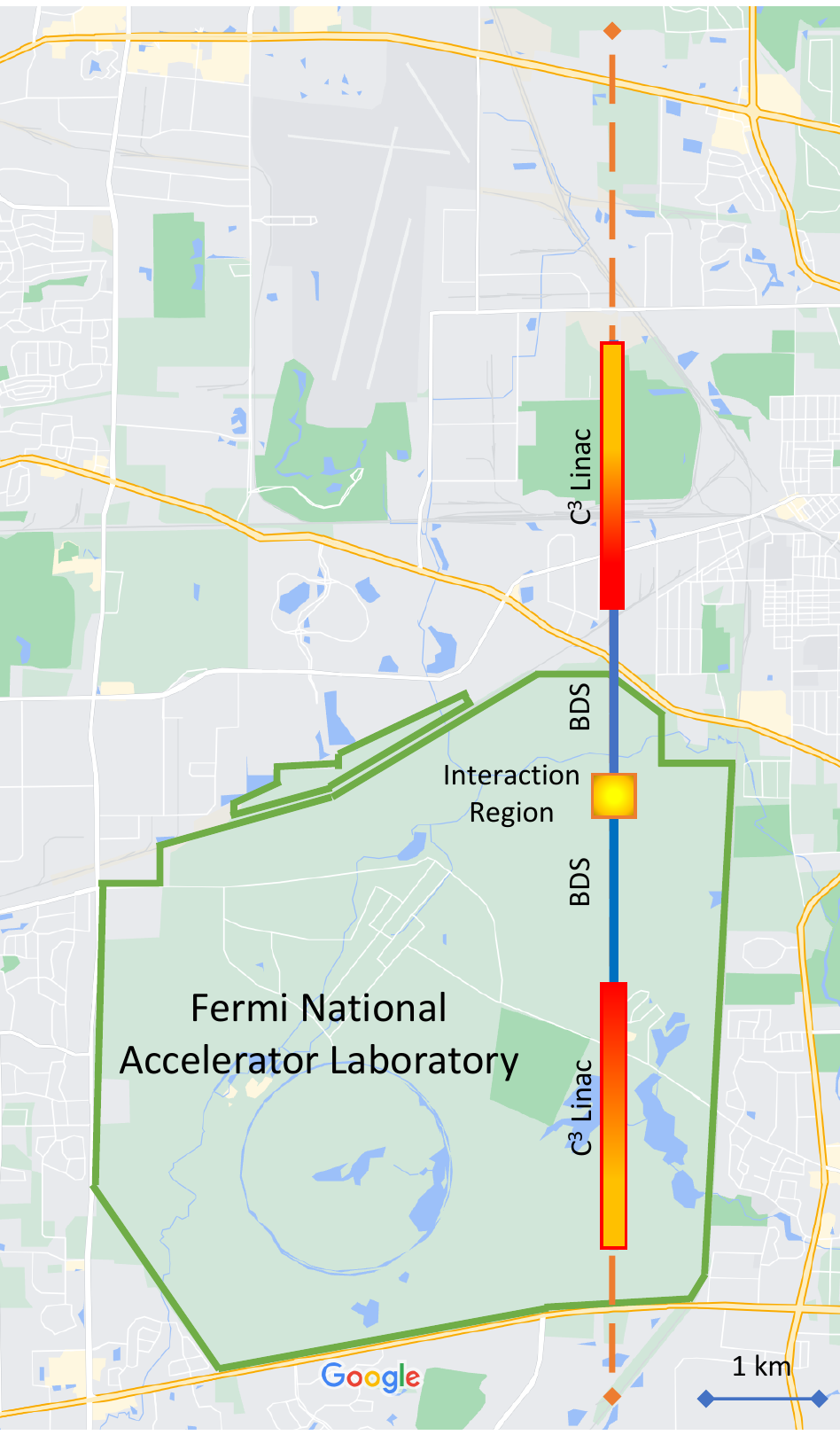}
    \caption{The 8-km footprint consisting of 5 km inside the Fermilab site and extending the facility under the Common Wealth Edison power company's easement.}
    \label{fig:LC_8kmLayout}
\end{figure}

For instance, the Fermilab site can fit a 7-km footprint linear machine entirely within its boundaries, (see Fig.~\ref{fig:LC_7kmLayout}) in North East-South West (NE-SW) orientation. The 8-km footprint currently proposed \CCC, reaching up to 550 GeV, can be accommodated at Fermilab, with about 5 km of the footprint inside the Lab site and extending the facility under the Common Wealth Edison power company's easement to the north of the Lab site - North-South (N-S) orientation.  It is also possible to further extend the footprint up to 12 km in this orientation, still keeping the interaction region of the collider within the Lab campus.  This siting location, shown in Fig. ~\ref{fig:LC_8kmLayout}, was, in fact, one of the options studied for the ILC at Fermilab.  Using the full 12 km length can provide upgrade paths to 750 GeV collision energy or higher. 

Perhaps, further optimization of the final focus could let the 8 km machine for energy upgrade up to 550 GeV fit within the boundary of the laboratory itself, using NE-SE orientation.  The details of these siting options are discussed in~\cite{FCGwhitepaper}.   

The initial exploration at the \CCC Higgs Factory could position the US to lead the drive to the TeV-scale, requiring larger machines. Sites such as the DOE Hanford site have room to accommodate even bigger footprint machines within their site boundary.

If the ILC goes forward in Japan, an energy upgrade using \CCC accelerators could be built, re-using the ILC damping rings, tunnel, and other conventional facilities. Such a machine could reach the TeV-scale as well.

\section{Detectors for \CCC Higgs Factory}

A detector for \CCC Higgs Factory must provide hermetic coverage with excellent charged particle momentum resolution, muon, electron and photon identification, and neutral particle energy resolution. The detector must have sufficient segmentation to enable excellent jet reconstruction capabilities, for example, using particle flow techniques. The tracking resolutions should enable high-precision reconstruction of the recoil mass in the ZH final state. It is anticipated that a material profile of less than 20\% X$_0$ is required for the tracker.  The first  sensor layer should be placed within 20 mm of the interaction point to allow very efficient $b$ and $c$ vertex tagging. These are general requirements applicable to any collider that will make precision measurements of the $\ee \rightarrow ZH$ process and other important Higgs factory reactions.  As such, there are already designs for ILC~\cite{White:2015dxj}, CEPC~\cite{CEPCStudyGroup:2018ghi}, CLIC~\cite{linssen2012physics,Robson:2018enq,CLICdp:2018vnx}, and FCC-ee~\cite{Bacchetta:2019fmz} that incorporate these requirements. We expect that all of these designs can be straightforwardly adapted for \CCC. Additionally, new detector technologies that have been developed in the LHC program could potentially be suitable for \CCC detector elements. Examples are MAPS technology for silicon tracking~\cite{Deptuch:2002tns}, highly segmented calorimetry~\cite{behnke2013international}, and dual readout calorimetry~\cite{Pezzotti:2022ndj}.  We anticipate that the experimentation at \CCC will attract a global community, and that a variety of proposals will be put forward. The eventual \CCC detector will arise from synergies or competition among these approaches.

An important difference among  \CCC, ILC, and circular colliders  that must be addressed is the different time structure of the bunch crossings.  Like ILC, \CCC delivers luminosity with trains of electron and positron bunches, with long gaps between them. This allows a design with power pulsing of the detector, lowering the requirements for cooling and making it easier to design a vertex and tracker detectors with a very low material budget~\cite{cccwhitepaper}.  The inter-train gaps at \CCC are shorter than those at ILC  (8.3 ms vs. 200 ms), but power pulsing requires only a gap of a few hundred $\mu$s to implement.  However, while ILC has relatively long time gaps between bunch collisions within a train (350-500 ns),  \CCC has gaps of only 3.5-5.5 ns.  For $\ee$ colliders at Higgs factory energies, there are underlying events, mainly from 1–2 soft $\gamma\gamma$ collisions to low-p$_T$ hadrons per bunch crossing. This is an unimportant background source for individual bunch crossings, but it might become significant if integrated over a bunch train.  Preliminary studies by the SiD collaboration show that this can be ameliorated by time-stamping of tracks to an accuracy of a few bunch crossings~\cite{ICHEPTimBtalk}.  Thus, we anticipate that the \CCC detector will be able to match the levels of precision projected for the ILC detectors from full-simulation analyses.  This issue should be studied, though, for the final \CCC detector designs.

Detectors that have been developed for future electron-positron collider experiments, such as ILD or SiD, can be adapted. As a concrete example, we have studied the adaptation of the SiD detector designed for the ILC to the requirements of \CCC.   The SiD concept is centered on a compact, cost-constrained design that addresses the full range of searches for new physics and precision measurements at a future high energy electron-positron linear collider. SiD is a silicon-based detector in a high magnetic field with excellent tracking and particle-flow calorimetry. Major design features are a robust silicon vertexing and tracking system, highly segmented calorimeters optimized for particle-flow, and a compact design with a 5T solenoid field. The SiD vertex detector consists of five barrel layers and forward/backward disks, with power pulsing and forced-air cooling. Requirements include a 3 micron (or less) hit spatial resolution, and a material budget of 0.1\% X$_{0}$. The combination of the 5T magnetic field and the envelope of the pair background allows, for ILC,  a first sensor layer at 14 mm from the beam. 

The main question about this design to be addressed for \CCC is the precise requirement for single bunch tagging and quantification of the needed time resolution. This implies that the pair background and occupancy studies done for ILC have to be repeated for the specific \CCC environment. Power-pulsing of SiD requires only a few hundred $\mu$s to settle after turning on.  Thus the current plan for forced-air cooling and power pulsing of the  the vertex detector and the main tracker should still be adequate for \CCC, though this needs to be confirmed.  Other areas that need to be revisited are the radius of the vertex detector inner layer, the beam pipe design, and the central magnetic field value, all in relation to the machine-specific pair background. In all, an SiD-like detector with a dedicated optimization of the vertex detector and other subsystems can be well adapted to the \CCC time structure to achieve the physics goals.

\section{Broader Impacts}
\subsection{Applications of Compact Linacs} 

R\&D on the \CCC concept is already being pursued at SLAC, UCLA, INFN, LANL and Radiabeam, along with closely related research in high gradient RF acceleration with CERN, KEK, PSI, MIT, and many other partners in the high gradient research community~\cite{HG2021}.   There is direct synergy with the development of compact electron accelerators for medical applications~\cite{Maxim2019,lu2021,Snively2020vhee,Nanni2020vhee}, high-flux x-ray sources for cargo screening applications \cite{weatherford2020modular} and the creation of compact X-ray free electron lasers~\cite{rosen2020,graves2020} (FELs).  The small size and low cost of compact X-ray sources based on \CCC technology will find a wide range of customers, including individual university research laboratories and major medical research centers.

\subsection{Workforce Development}

The particle physics community has traditionally attracted brilliant students from the incoming classes in most university physics departments. The opportunity to work on fundamental problems while using state-of-the-art technologies, and develop them as necessary, is the leading motivator for the students. The \CCC will provide much needed impetus to make particle physics programs attractive to the young and future generations. Due to the long lead times involved in design, construction and operation of the programs, it is important to involve the new generation entering graduate classes early in the design process. Many of those early students who will develop expertise in compact accelerator development or new detector technologies will build the \CCC and conduct research using it, while others will contribute to the society at large. 

\subsection{Future HEP Facilities}

The \CCC facility would represent a significant investment in the future of the Energy Frontier. Its design must allow for the extension of the facility's reach in energy as we are guided by the physics we discover during the operation of the facility. Indeed, this is inherent to the \CCC concept even in its initial stages through the facility's approach for scaling from 250 GeV to 550 GeV center of mass energy. While one possibility for extension in energy is the use of rf accelerator technology, either with a larger facility footprint or through the development of a higher gradient main linac, we should also consider the adoption of advanced accelerator technology for a future upgrade. During the timeline of the \CCC R\&D demonstration plan and the eventual preparation of a \CCC TDR, advanced and novel accelerators will be working towards furthering their own technical milestones for laser and beam driven plasma wakefield and structure wakefield \cite{LPA,PWFA,SWA}.  These studies may be advanced enough to inform future requirements on the facility (for example tunnel diameter, configuration of the BDS, distribution of power and cooling) that may be minimally invasive during initial construction, but prohibitively expensive for a future upgrade. Keeping this potential avenue in mind and open may save significant infrastructure and construction costs for future upgrades.

\section{Towards a \CCC Conceptual Design}

Developing a conceptual design with a defensible cost estimate for a \CCC Higgs Factory is a requirement for proposing a project to the US funding agencies and attracting worldwide community interested in Higgs studies at \CCC. The R\&D required to prepare a conceptual design for \CCC includes the following elements in a proposed demo facility~\cite{CCCdemo}, that will provide a full demonstration of the 
\CCC Main Linac technology on the GeV scale. 

    
The outstanding technical achievements of the ILC, CLIC, and NLC collaborations are central to the rapid realization of the \CCC proposal. Many of the subsystems for the accelerator complex are interchangeable between linear collider concepts with manageable modifications to account for variations in pulse format and beam energy.  Because these subsystem designs are already mature, the \CCC demonstration facility can  focus on the specific  set of technical milestones associated with the \CCC concept itself:

\begin{itemize}
\item Development of a fully engineered and operational cryomodule including linac supports, alignment, magnets, BPMs,  RF/electrical feedthroughs, liquid and gaseous nitrogen flow, and safety features.
\item Operation of the cryomodule under the full thermal load of the Main Linac and maximum liquid nitrogen flow velocity over the accelerators in the cryomodule, demonstrating an  acceptable level of  vibrations.
\item Operation with a multi-bunch photo-injector to induce wake fields, using high charge bunches 
 and a tunable-delay witness bunch.
\item Achievement of 120~MeV/m accelerating gradient in single bunch mode for an energy gain of 1~GeV in a single cryomodule, including tests at higher gradients to establish breakdown rates.
\item Acceleration and wakefield effect measurements with a fully damped-detuned structure.
\item Development, in partnership with industry, of the baseline  C-band RF source unit that will be installed with the Main Linac. The RF source unit will be modified from existing industrial product lines.

\end{itemize}



\section{Outlook}

With a five-year \CCC demonstrator project (described in ~\cite{CCCdemo}), the project will be ready for US DOE project review processes, on the way to attracting international participation in the \CCC facility. 

The timeline for the developing the CDR is driven by the technical timeline for maturing the main linac technology as part of the \CCC demonstration. Not all objectives of the demonstration are required to develop the CDR. However, early stage technology maturation and demonstrations would allow for the necessary detail of the CDR.  The CDR would also rely heavily on the complimentary efforts of the ILC and CLIC community which have advanced the readiness of the injector complex, damping rings, and BDS.

The \CCC demonstrator project will involve multiple national laboratories and some university groups. The demonstrator itself could revitalize the US HEP community, attracting bright students and postdoctoral fellows, especially those graduating from the LHC project, which is underway at CERN. The commercialization of the accelerating structures and RF sources, through the demonstrator project, will enable long term partnerships with industry to build the \CCC and also build a new market for small-footprint X-ray sources.

The signatories of this white paper strongly believe that our community strategy for understanding the Higgs physics is significantly enhanced by the development of the \CCC project. We urge colleagues participating in the Snowmass community studies to strongly support the \CCC demonstrator project, with the aim of preparing a conceptual design report within five years.

\section{Acknowledgements}

The work of Tim Barklow, Martin Breidenbach, Michael Kagan, Emilio A. Nanni, Michael E. Peskin, Ariel G. Schwartzman, Su Dong, Caterina Vernieri is supported  by Department of Energy Contract DE-AC02-76SF00515. The work of Jim Brau is supported by Department of Energy  DE-SC0017996.

\bibliographystyle{atlasnote}
\bibliography{bibliography.bib}

\end{document}